\documentclass[article,screen,nonacm]{acmart}

\usepackage[utf8]{inputenc} 
\usepackage[T1]{fontenc}    
\usepackage{hyperref}       
\usepackage{url}            
\usepackage{booktabs}       
\usepackage{amsfonts}       
\usepackage{nicefrac}       
\usepackage{microtype}      
\usepackage{lipsum}		
\usepackage{graphicx}
\usepackage{natbib}
\usepackage{doi}

\usepackage{url,hyperref,lineno}
\usepackage{subfig}
\usepackage[T1]{fontenc}

\title{Application of the Cyberinfrastructure Production Function Model to R1 Institutions}

\begin{document}

\title{Application of the Cyberinfrastructure Production Function Model to R1 Institutions}

\author{Preston M. Smith}
\email{psmith@purdue.edu}
\affiliation{%
  \institution{Purdue University}
  \city{West Lafayette, IN}
  \country{USA}
}

\author{Jill Gemmill}
\email{gemmill@clemson.edu}
\affiliation{%
  \institution{Clemson University}
    \city{CLemson, SC}
  \country{USA}
}

\author{David Y. Hancock}
\email{dyhancoc@iu.edu}
\affiliation{%
  \institution{Indiana University}
    \city{Bloomington, IN}
  \country{USA}
}

\author{Brian W. O'Shea}
\email{wsnappch@iu.edu}
\affiliation{%
  \institution{Michigan State University}
    \city{East Lansing, MI}
  \country{USA}
}

\author{Winona Snapp-Childs}
\email{wsnappch@iu.edu}
\affiliation{%
  \institution{Indiana University}
    \city{Bloomington, IN}
  \country{USA}
}

\author{James Wilgenbusch}
\email{jwilgenb@umn.edu}
\affiliation{%
  \institution{University of Minnesota}
    \city{St. Paul, MN}
  \country{USA}
}
\authorsaddresses{}

\begin{abstract}
High-performance computing (HPC) is widely used in higher education for modeling, simulation, and AI applications. A critical piece of infrastructure with which to secure funding, attract and retain faculty, and teach students, supercomputers come with high capital and operating costs that must be considered against other competing priorities. This study applies the concepts of the production function model from economics with two thrusts: 1) to evaluate if previous research on building a model for quantifying the value of investment in research computing is generalizable to a wider set of universities, and 2) to define a model
with which to capacity plan HPC investment, based on institutional production - inverting the production function. We show that the production function model does appear to generalize, showing positive institutional returns from the investment in computing resources and staff.  We do, however, find that the relative relationships between model inputs and outputs vary across institutions, which can often be attributed to understandable institution-specific factors.
 
\end{abstract}

\maketitle
\renewcommand{\shortauthors}{Smith et al.}

\section{Introduction}

 In higher education, university leaders must balance competing priorities in which to invest funding for infrastructure. While federal funding opportunities do exist for cyberinfrastructure, it is best treated as a coordinated investment by many stakeholders, including the universities themselves.
 
 At a given university, a vice president for research must decide if a proposed new electron microscope adds the most value, or is it a wind tunnel instead? Even within an IT organization, a CIO must choose between research IT, the enterprise resource planning system, student wireless internet, and more. In order for research computing and data to reach the top of any of these lists, a clear alignment of the value proposition to the institution's goals and measures of success must be communicated.

However, many research computing organizations are immature at communicating their value, instead approaching the importance of investing in research computing as an article of faith. \cite{Ludwig2012} described the state of the community's quantification of costs, benefits, and their relationship as ``There is not much research available to answer these questions. In fact: \textbf{almost no research}'' 

This claim by Ludwig is further illustrated by  \cite{10.1145/3437359.3465589}, where the authors noted that relatively few centers even measure the usage of their systems (41 of 69), academic outcomes (33 of 69), and ROI (32 of 69). 

This study's goals are two-fold: 
\begin{enumerate}
    \item Apply the production model for cyberinfrastructure described by \cite{SmithMetrics2024} to several additional institutions to evaluate the model's suitability for general use.
    \item Based on surveyed real-world data, invert the production function to create a model for sizing cyberinfrastructure investment based on levels of output.
\end{enumerate}

\section{Materials and Methods}

\subsection{Literature Review}

A systematic literature review of ROI from educational research grant funding can be found in \cite{rahman2252return} where the authors examine published ROI studies over a twenty year period from 2001 to 2021 and sort deliverables into output and outcomes, specifically those that are tangible and intangible. They also discuss the different types of return on investment measurement found in those studies: financial, cost benefit analysis, return on equity, and social return on investment. A case is made to define and communicate research grant output definitions early so that uniform research evaluations and measurement of the impact and ROI of research funding can be accomplished.

There are also macro-analysis studies of ROI in research infrastructures as highlighted by \cite{DELBO201626} in an analysis of research infrastructure investment in the European Union between 1987 and 1991. Based on prior publication the author concludes the primary method of analysis for specific projects should use a cost benefit analysis approach.The author introduces a new cost effectiveness measure that factors in publications to provide a proxy rate of return and notes the difference in definition of similar terms in macro and micro studies. The conclusion at the macro-level is a positive contribution to economic growth and productivity from investment in large research infrastructures and notes the open issues of time lag over which the activities exert an impact, in addition to spillover between geographical regions and industries. 

More detailed reviews of literature on measuring the value of cyberinfrastructure that serves as the theoretical foundation for this work may be found in \cite{SmithMetrics2024} and \cite{10.1145/3491418.3535131}. The following sections summarize the key outcomes from these previous works. 

\subsubsection{ROI in Cyberinfrastructure}

The methods utilized in this study are described in \cite{SmithMetrics2024}, which employed the concept of the production function \citep{cobbdouglas} from economics as a tool to describe how CI investment contributes to institutional output. Apon's works (\citeyear{Apon2010}, \citeyear{Apon2015}) were some of the first studies of the ROI of cyberinfrastructure, and their exploration of how the presence of a robust cyberinfrastructure impacted institutional and departmental productivity were a key influence for Smith's model.

At Indiana University, \cite{Snapp-ChildsMetrics2024} used the International Integrated Reporting Framework to assess the cost of the XSEDE cyberinfrastructure and the value of the end products created using it.   \cite{Stewart2019}, coining the notion of ROI\textsubscript{proxy}, noted that ``A value of greater than one indicates that what was done would have cost more if done by an alternative method.''  

\cite{Stewart} found positive ROI (greater than 1) for each of a campus supercomputer, Jetstream, and XSEDE  by assigning financial values to them by comparing to commercial services or user surveys.  

\cite{SmithMetrics2024} developed his model to build upon Apon's, specifically adding a people dimension and addressing the decreasing utility of the Top 500 list as a proxy for an institution's computing capacity. Smith only assessed investment and outputs of one institution, however, leading to a clear opportunity for improvement by applying the production function model to more institutions than just Purdue. It is this gap that this study seeks to fill.

Several articles provide a theoretical base on selecting metrics with which to measure CI investment and institutional output.  \cite{Stewart2019}
 described several potential outcome metrics, including the amount of grant income, new products and patents, the amount of economic impact, and new research innovations (e.g., papers, citations).

 \cite{Arora2024} ``developed a taxonomy of the CI projects and defined a set of standard metrics for the assessment of the projects belonging to the different categories in the taxonomy.'' Various metric categories evaluated include outcomes like publications, citations, IP, time to degree, and the resulting MICI (Measuring the Impact of CyberInfrastructure) model introduces an impact score as a broadly applicable way to measure CI projects.

 \cite{/content/paper/3ffee43b-en} proposed a ``Framework for assessing the scientific and socio-economic impact of Research Infrastructures'' and linked 58 potential impact indicators for research infrastructure (RI) to strategic objectives. These indicators include the number of publications, citations, high impact publications, grants, patents, students trained, and more. 

 In a 2014 workshop report, \citet{Berente2014} identified four dimensions for CI value outcomes: economic, science, workforce, and innovation. In a subsequent panel at SC14 \citep{Newby2014}, speakers described input investments as ``physical infrastructure, energy, capital equipment and software and personnel''. Academic products (outputs) are described as ``Intellectual/Reputation ROI (publications, graduations, numbers of faculty and student users, and enhancement to reputation); and Financial ROI (grants and contracts, recharge fees, and patents/royalties).''

\subsubsection{The Production Function Model}
\label{sec:prod_func_lit}

Charles Cobb and Paul Douglas' 1928 article ``A Theory of Production'' \citep{cobbdouglas}, described a statistical model that framed the output of production \emph{Y}, using the inputs of capital \emph{C}, and of labor \emph{L}. 

The production function model is represented in equation form in Equation \ref{eq:cobb_douglas}. 
\begin{equation} \label{eq:cobb_douglas}
   Y = f(L,C) 
\end{equation} 

Using a least-squares regression, Cobb and Douglas reported that the production function had coefficients of 0.75 for \emph{L} and 0.25 for \emph{C}, indicating a 75\% and 25\% proportion of labor and capital's contribution to the output of the function.

Comparing to Cobb and Douglas' 75:25 proportion in the academic cyberinfrastructure sector, \cite{SmithMetrics2024} reported production function labor and capital coefficients for Purdue University of 63:31, 65:39, 60:35, or 58:36 respectively, depending on the output being modeled.

\subsection{Historical Dataset}
\label{sec:hist_dataset}

A dataset was collected from five participating R1 institutions between 2011 and 20223 that included the  annual year-over-year HPC capacity centrally operated and the annual university-funded salary and wage expenses relating to that centralized HPC. These self-reported, internal data are combined with publicly available metrics of outputs: R\&D expenditures as reported to the NSF HERD Survey, publications, the high-impact subset of those publications tracked in the Nature Index, and earned doctorates. 

Input and output metrics are summarized in Table \ref{tab:datalist}, and further details about metrics are further described in Section \ref{sec:inputlist} and \ref{sec:inputlist}.

\begin{table}[ht]
\centering
  \caption{Selected Input and Output Metrics}
\begin{tabular}{l|l}
  \hline
 Input & Output  \\ 
  \hline
Labor: Annual salary costs & R\&D expenditures reported to HERD \\
Capital: Total TF operated &  Earned doctorates reported to SED \\
                  &  Total publications produced, per Scopus \\
                  &  High-impact ``Nature Index'' publications 
\end{tabular}
  \label{tab:datalist}%
\end{table}

\subsubsection{Input Metrics}
\label{sec:inputlist}

The  set of annual input and output metrics is described in further detail below:
\begin{itemize}
    \item \textbf{TeraFLOPS of compute capacity available}: This metric represents the aggregate compute capacity centrally operated at the institution during a given year. A proxy for overall center size, this information was assembled from internal records maintained by center leadership. These values do not include HPC resources that are funded and operated on behalf of outside entities, or lab/department scale resources. 
    \item \textbf{Total salary costs for the HPC center}: This metric is the total salary and wage expenses for the campus Research Computing and Data (RCD) organization per internal financial systems. These dollars do not include expenditures on grant accounts, as this study is investigating \emph{institutional} investment. 
\end{itemize}

Capacity in FLOPS is used rather than a raw count of cores, nodes, or GPUs to account for the relative performance of those cores or GPUs as time goes on. For example:  10,000 7-year old cores are likely not an indicator of greater capacity or investment compared to 7,000 1-year old cores. The 1 year-old cores will most likely have a greater peak performance than the 7-year old system. Further, a center with large but extremely old resources may actually be one subject to under-investment. For labor metrics, FTE count is potentially an equivalent proxy for investment, but potentially more difficult to track historically than salary dollars, which can be reported from institutional financial systems.

\subsubsection{Output Metrics}
\label{sec:outputlist}
The following annual data are described in further detail below:
\begin{itemize}
    \item \textbf{HERD-reported R\&D Expenditures}: This metric represents the total annual institutional R\&D expenditures reported to the NSF Higher Education Research and Development (HERD) survey \footnote{https://www.nsf.gov/statistics/srvyherd}. 
    \item \textbf{Earned Doctorate Degrees}: This metric represents the number of Doctoral degrees awarded by each institution annually, reported to the NSF Survey of Earned Doctorates (SED) \footnote{https://www.nsf.gov/statistics/doctorates/}.
    \item \textbf{Total Publications by authors affiliated with the institution}: This metric is the total number of publications reported in Scopus \footnote{https://scopus.com}, for a given year, with an affiliation tying to the institution
        \item \textbf{Total Publications in High-Impact Journals}: This metric is the total number of publications reported in Scopus for a given year, where an author with the institution's affiliation published in one of the high-quality journals listed in the Nature Index \footnote{https://www.natureindex.com/} .
\end{itemize}




\subsection{Current Investment Dataset}
In Spring 2025, a survey was sent to Coalition for Academic Scientific Computation (CASC) member institutions and members of the Midwest Research Computing and Data Consortium (MWRCD) to obtain a broader dataset of system capacity and salary investment to generate a broad dataset with which to develop a benchmarking model, and create a base for creating a future broader-scale historical dataset, to address concerns with the small size of the historical datasets. 28 responses were received, all from R1 Carnegie Classification universities, with R\&D expenditures ranging from under \$180M to \$2B.

These survey responses were merged with the most recent (2023) HERD and Survey of Earned Doctorates (SED) data for each institution, and the data were cleaned. Some respondents were unable to provide capacity in TeraFLOPS, and we worked with the respondent to calculate a value based on their reported core and GPU counts by their hardware specifications. Similarly, some respondents were unable to report a dollar amount for staff costs but provided an FTE count. In this case, the FTE count was multiplied by the median RCD professional compensation of \$90k USD reported by \cite{10.1145/3569951.3593599}.

Each institution's ratios of TeraFLOPS and salary costs to 2023 HERD expenditures were calculated, and the ratios then averaged to find a coefficient with which to describe system capacity and salary investment as a function of R\&D expenditures.

R\&D expenditures is not the only driver of computing demand. Another way to quantify the need for capacity at a research institution is in the most important product of all - its graduates. An institution that graduates high numbers of STEM PhDs will have a greater need for a well-resourced cyberinfrastructure than one that only graduates a few. 

To model this additional dimension,  each institution's ratios of TeraFLOPS and salary costs to 2023 earned doctorates were calculated. As with expenditures, these relationships were then averaged to identify a coefficient with which to size system capacity and salary investment as a function of PhDs awarded.

\subsection{Correlation Analysis}
Next, with the historical dataset, we will perform a correlation analysis of single institutional research computing investment factors with university outputs.

\textbf{Hypothesis:} there is universally a strong correlation between institutional inputs and outputs, as has been shown at Purdue University in \cite{SmithMetrics2024}. 

\subsection{Production Function Model}
Following correlation analysis, the production function regression models will be performed first against the combined historical dataset and then individually for each institution, regressing institutional outputs on compute capacity and RCD staff investment. Regression coefficients will be summarized, compared to those for Purdue identified by \cite{SmithMetrics2024}, and the coefficients translated to real-world impact on the selected output.

\textbf{Hypothesis:} the combined and individual production function models will be highly-performing for all institutions examined, as with Purdue in \cite{SmithMetrics2024}.

\subsection{Relative Importance of Coefficients}

Using the \texttt{lmg} method implemented in the R package \texttt{relaimpo} \citep{Gromping2007}, we will calculate the proportion of the overall variance explained by each input in the model. 

\textbf{Hypothesis:} the proportions of variance on the combined and individual  institutions' production function models will be similar to Purdue's in \cite{SmithMetrics2024}.


\subsection{Benchmark: Investment as a function of outputs}

\textbf{Research question:} Can the existing relationships between a campus' RCD investment and its outputs be generalized to create a best practice to guide decision makers operating a university HPC center?

A question at the front of the mind of any research computing leader is ``is my center appropriately resourced for the expectations of researchers at my university?'' Comparing raw numbers to each other does not provide a good benchmark - a \$300M/year research enterprise is unlikely to require the level of resources that a billion-dollar university does. Much like nations measure defense or healthcare costs as a fraction of GDP, we will measure RCD investment as a function of key outputs.

Specifically, with the benchmarking dataset we will answer these questions:
\begin{itemize}
\item \textbf{Benchmark Q1:} What are the observed ratios of computing capacity to outputs? 
\item \textbf{Benchmark Q2:} What are the observed percentage of salary costs vs. outputs?
\item \textbf{Benchmark Q3:} With this information, can we present formulae for appropriately sizing an RCD center's people and capital resources?

\end{itemize}

\section{Results}

\subsection{Correlation Analysis}
\label{sec:corr_anal}
First, we perform a correlation analysis using the Kendall Tau rank correlation, comparing the capital input TeraFLOPS and labor input salaries to identified outputs. Kendall correlation is used rather than Pearson due to not only to the relatively small dataset sizes per institution, but also the non-normal distribution of TeraFLOPs and salaries for each institution. For TeraFLOPs, all institutions show a strong correlation with each output, with the exception of Institution D's earned doctorate output.

\begin{table}[htbp]
  \centering

    \begin{tabular}{lrrrrr}
          & \multicolumn{1}{l}{Institution A} & \multicolumn{1}{l}{Institution B} & \multicolumn{1}{l}{Institution C} & \multicolumn{1}{l}{Institution D} & \multicolumn{1}{l}{Institution E} \\
    HERD Expenditures & 0.670 & 0.948 & 0.868 & 0.901 & 0.930 \\
    Publications & 0.961 & 0.958 & 0.898 & 0.878 & 0.918 \\
    Hi Impact Pubs & 0.585 & 0.886 & 0.853 & 0.795 & 0.916 \\
    Earned Doctorates & 0.716 & 0.599 & 0.515 & 0.219 & 0.870 \\
    \end{tabular}%
    \caption{Correlation Coefficients - TF vs Outputs, by Institution}
  \label{tab:tfcorr}%
\end{table}%

Salary investment from all institutions shows a moderate to strong correlation with most outputs ($\simeq 0.6$ to 0.9), with Institution A's salary correlations being the weakest overall.  Institution D's correlation with earned doctorates is not meaningful (it's negative, here, in fact), due to external institutional trends in doctoral degrees awarded discussed in greater detail in Section \ref{sec:localfactors}.

\begin{table}[htbp]
  \centering
  
    \begin{tabular}{lrrrrr}
          & \multicolumn{1}{l}{Institution A} & \multicolumn{1}{l}{Institution B} & \multicolumn{1}{l}{Institution C} & \multicolumn{1}{l}{Institution D} & \multicolumn{1}{l}{Institution E} \\
    HERD Expenditures & 0.295 & 0.895 & 0.706 & 0.857 & 0.767 \\
    Publications & 0.543 & 0.876 & 0.647 & 0.786 & 0.751 \\
    Hi Impact Pubs & 0.421 & 0.838 & 0.647 & 0.429 & 0.767 \\
    Earned Doctorates & 0.490 & 0.562 & 0.657 & -0.429 & 0.660 \\
    \end{tabular}%
  \caption{Correlation Coefficients - Salary Costs vs Outputs, by Institution}
  \label{tab:salcorr}%
\end{table}%

\subsection{Production Function Model}
\label{sec:prodfunc}

Table \ref{table:comb_regs} reports the results of the and  of the production function model across the entire multi-year combined dataset (n=82) of all five studied institutions. The resulting models report moderate adjusted R\textsuperscript{2} values for each output, with statistically significant predictors. 

Despite only moderately high adjusted R\textsuperscript{2} and a limited sample size of institutions, this application of the production model to this population shows promise as first attempt to replicate \cite{SmithMetrics2024} with a model incorporating more than a single institution. Table \ref{tab:combsummary} reports the results of the combined production function model in plain terms, with, for example, each increase of 100 TeraFLOPs corresponding with an increase of \$2.99 million in annual HERD R\&D Expenditures. Each \$100k investment in RCD salaries corresponds with an increase of \$14.46 million of annual HERD expenditures.

\begin{table}[h]
\begin{center}
\begin{tabular}{l c c c c}
\hline
 & Publications & Earned Doctorates & HERD Expenditures & High Impact Publications \\
\hline
(Intercept) & $1934.814^{***}$ & $301.258^{***}$ & $154.743^{***}$ & $254.473^{***}$ \\
            & $(409.760)$      & $(35.628)$      & $(38.128)$      & $(62.676)$      \\
TF          & $0.196^{*}$      & $0.012$         & $0.030^{***}$   & $0.036^{*}$     \\
            & $(0.090)$        & $(0.008)$       & $(0.008)$       & $(0.014)$       \\
Salaries    & $1369.031^{***}$ & $78.246^{***}$  & $144.637^{***}$ & $212.089^{***}$ \\
            & $(189.689)$      & $(16.493)$      & $(17.650)$      & $(29.014)$      \\
\hline
R$^2$       & $0.535$          & $0.341$         & $0.634$         & $0.557$         \\
Adj. R$^2$  & $0.524$          & $0.325$         & $0.625$         & $0.546$         \\
Num. obs.   & $86$             & $86$            & $86$            & $86$            \\
\hline
\multicolumn{5}{l}{\scriptsize{$^{***}p<0.001$; $^{**}p<0.01$; $^{*}p<0.05$}}
\end{tabular}
\caption{Production Function Models - All Institutions}
\label{table:comb_regs}
\end{center}
\end{table}

\begin{table}[htbp]
  \centering

    \begin{tabular}{lcccc}
          & \multicolumn{1}{p{6.335em}}{\textbf{Publications}} & \multicolumn{1}{p{6em}}{\textbf{Earned Doctorates}} & \multicolumn{1}{p{5.75em}}{\textbf{HERD Expenditures}} & \multicolumn{1}{p{5.665em}}{\textbf{High Impact Pubs}} \\
    \textbf{100 TeraFLOPS} & 19.59 & 1.24  & 2.99  & 3.59 \\
    \textbf{\$100k Salaries} & 136.90 & 7.83  & 14.46 & 21.21 \\
    \textbf{Adj. R$^2$} & .524 & .325 & .625 & .546 \\
    \end{tabular}%
      \caption{Production Function Models - Combined Dataset}
  \label{tab:combsummary}%
\end{table}%

As we explore deeper, we apply the production function model to each institution individually, as performed by \cite{SmithMetrics2024}. Each subsection in Section \ref{sec:prodfunc} examines an output (Publications, Earned Doctorates, High-Impact Publications, and HERD R\&D Expenditures) and presents each individual institution's model results.

\subsubsection{Publications}
\label{sec:pub_prod}
In Table \ref{table:pub_coefficients}, production function models on publications exhibit high adjusted R\textsuperscript{2} for all institutions, 3 of which have both coefficients exhibiting a a p-value \textless 0.05.  The high p-values for Institution D is likely due to the limited number of observations (only 8 years), and we anticipate that model predictors would be statistically significant with the addition of more historical data points. Institution A's salary coefficient is low and non-significant, due to historical volatility in investment levels. (Section \ref{sec:localfactors})

\begin{table}[h]
\begin{center}
\begin{tabular}{l c c c c c}
\hline
 & Institution A & Institution B & Institution C & Institution D & Institution E \\
\hline
(Intercept) & $1508.149^{***}$ & $1521.613^{***}$ & $4340.957^{***}$ & $9229.195^{***}$ & $1987.484^{***}$ \\
            & $(112.468)$      & $(161.583)$      & $(171.112)$      & $(1077.371)$     & $(250.412)$      \\
TF          & $0.404^{***}$    & $0.082^{***}$    & $0.216^{***}$    & $0.231$          & $0.231^{***}$    \\
            & $(0.045)$        & $(0.018)$        & $(0.046)$        & $(0.098)$        & $(0.043)$        \\
Salaries    & $230.963^{*}$    & $1135.829^{***}$ & $870.411^{***}$  & $149.197$        & $1340.362^{***}$ \\
            & $(88.555)$       & $(58.961)$       & $(132.982)$      & $(332.278)$      & $(123.793)$      \\
\hline
R$^2$       & $0.918$          & $0.977$          & $0.921$          & $0.784$          & $0.938$          \\
Adj. R$^2$  & $0.904$          & $0.974$          & $0.910$          & $0.697$          & $0.933$          \\
Num. obs.   & $15$             & $21$             & $17$             & $8$              & $25$             \\
\hline
\multicolumn{6}{l}{\scriptsize{$^{***}p<0.001$; $^{**}p<0.01$; $^{*}p<0.05$}}
\end{tabular}
\caption{Production Function Models - Predicting Publications}
\label{table:pub_coefficients}
\end{center}
\end{table}

Table \ref{tab:pubsummary} summarizes the results of the production function model on publications in real-world terms. For Institution C, each increase of 100 TeraFLOPs operated corresponds with an annual increase of 21.57 publications coming from Institution C, and each \$100k investment in RCD salaries corresponds with an annual increase of 87.04 publications.

\begin{table}[htbp]
  \centering

    \begin{tabular}{lccccc}
 & Institution A & Institution B & Institution C & Institution D & Institution E \\
    \textbf{100 TeraFLOPS} & 40.41 & 8.18  & 21.57 & 23.09 & 23.12 \\
    \textbf{\$100k Salaries} & 0.04 & 113.60 & 87.04 & 14.92 & 134.00 \\
    \textbf{Adj. R$^2$} & .904 & .974 & .921 & .697 & .933 \\
    \end{tabular}%
  
    \caption{Summary - Publications Output}
    \label{tab:pubsummary}%
\end{table}%

\subsubsection{Earned Doctorates}
Table \ref{table:phd_coefficients} describes the results of production function models on earned doctorates. For this output, these institution-level models do not exhibit the same significant relationship as found in \cite{SmithMetrics2024}. 

Institution A, Institution B, and Institution C have only one significant predictor.  For Institution D and Institution A in particular, the model fails to show significance, due to the external factors related to the output itself discussed in Section \ref{sec:localfactors}.

\begin{table}[h]
\begin{center}
\begin{tabular}{l c c c c c}
\hline
 & Institution A & Institution B & Institution C & Institution D & Institution E \\
\hline
(Intercept) & $168.355^{***}$ & $308.485^{***}$ & $439.563^{***}$ & $622.061^{*}$ & $380.944^{***}$ \\
            & $(14.787)$      & $(24.282)$      & $(11.909)$      & $(202.725)$   & $(19.239)$      \\
TF          & $0.026^{***}$   & $-0.001$        & $-0.003$        & $-0.020$      & $0.021^{***}$   \\
            & $(0.006)$       & $(0.003)$       & $(0.003)$       & $(0.018)$     & $(0.003)$       \\
Salaries    & $23.798$        & $47.405^{***}$  & $38.816^{***}$  & $30.359$      & $98.548^{***}$  \\
            & $(11.643)$      & $(8.860)$       & $(9.255)$       & $(62.524)$    & $(9.511)$       \\
\hline
R$^2$       & $0.767$         & $0.698$         & $0.618$         & $0.238$       & $0.941$         \\
Adj. R$^2$  & $0.728$         & $0.665$         & $0.563$         & $-0.067$      & $0.936$         \\
Num. obs.   & $15$            & $21$            & $17$            & $8$           & $25$            \\
\hline
\multicolumn{6}{l}{\scriptsize{$^{***}p<0.001$; $^{**}p<0.01$; $^{*}p<0.05$}}
\end{tabular}
\caption{Production Function Models - Predicting Earned Doctorates}
\label{table:phd_coefficients}
\end{center}
\end{table}

Table \ref{tab:phdsummary} summarizes the results of the production function model on earned doctorates in real-world terms. For example, we can see that for Institution E, each additional 100 TeraFLOPs aligns with with an annual increase of 2.10 PhDs awarded , and each \$100k investment in RCD salaries corresponds with an additional 9.86 earned doctorates annually. No other institution's model provide a meaningful result.

\begin{table}[htbp]
  \centering

    \begin{tabular}{lccccc}
 & Institution A & Institution B & Institution C & Institution D & Institution E \\
    \textbf{100 TeraFLOPS} & 2.62  & -0.05 & -0.26 & -2.05 & 2.10 \\
    \textbf{\$100k Salaries} & 2.38  & 4.74  & 3.88  & 3.04 & 9.86 \\
    \textbf{Adj. R$^2$} & .728 & .665 & .563 & -.067 & .936 \\
    \end{tabular}%
    \caption{Summary - Earned Doctorates Output}
   \label{tab:phdsummary}%
\end{table}%

\subsubsection{HERD Expenditures}

For HERD R\&D Expenditures, all institutions exhibit a strong production function model with a high adjusted R\textsuperscript{2}, indicating that using the production function as a model appears to broadly hold true. Only Institution A has an insignificant predictor (salaries) atypical from the other institutions, the small coefficient further reinforcing that the volatility in Institution A's investment/disinvestment cycle make the salaries variable meaningless for the purpose of modeling. (Section \ref{sec:localfactors})

\begin{table}[h!]
\begin{center}
\begin{tabular}{l c c c c c}
\hline
 & Institution A & Institution B & Institution C & Institution D & Institution E \\
\hline
(Intercept) & $157.890^{***}$ & $122.298^{*}$   & $374.975^{***}$ & $440.168^{***}$ & $184.293^{***}$ \\
            & $(17.901)$      & $(46.438)$      & $(23.840)$      & $(58.239)$      & $(26.299)$      \\
TF          & $0.046^{***}$   & $0.021^{***}$   & $0.034^{***}$   & $0.027^{**}$    & $0.032^{***}$   \\
            & $(0.007)$       & $(0.005)$       & $(0.006)$       & $(0.005)$       & $(0.005)$       \\
Salaries    & $3.075$         & $130.964^{***}$ & $112.268^{***}$ & $136.264^{***}$ & $118.495^{***}$ \\
            & $(14.095)$      & $(16.945)$      & $(18.528)$      & $(17.962)$      & $(13.001)$      \\
\hline
R$^2$       & $0.817$         & $0.906$         & $0.922$         & $0.986$         & $0.937$         \\
Adj. R$^2$  & $0.787$         & $0.895$         & $0.911$         & $0.981$         & $0.931$         \\
Num. obs.   & $15$            & $21$            & $17$            & $8$             & $25$            \\
\hline
\multicolumn{6}{l}{\scriptsize{$^{***}p<0.001$; $^{**}p<0.01$; $^{*}p<0.05$}}
\end{tabular}
\caption{Production Function Models - Predicting HERD Expenditures}
\label{table:herd_coefficients}
\end{center}
\end{table}

Table \ref{tab:herdsummary} summarizes the results of the production function model on HERD Expenditures in real-world terms. For example, we can see that for Institution B each additional 100 TeraFLOPs represents an annual increase of \$2.13M of research expenditures, and each \$100k investment in RCD salaries corresponds with an additional \$13.10M of annual expenditures.

\begin{table}[h]
  \centering

    \begin{tabular}{lccccc}
 & Institution A & Institution B & Institution C & Institution D & Institution E \\
    \textbf{100 TeraFLOPS} & 4.64  & 2.13  & 3.38  & 2.73  & 3.18 \\
    \textbf{\$100k Salaries} & 0.31 & 13.10 & 11.23 & 13.63 & 11.85 \\
    \textbf{Adj. R$^2$} & .787 & .895 & .911 & .981 & .931 \\
    \end{tabular}%
    \caption{Summary - HERD Expenditures Output}
  \label{tab:herdsummary}%
\end{table}%

\subsubsection{High-Impact Publications}
The production function models for high-impact publications are varied - Institution E and Institution C both have models that explain significant amounts of the output, with both a high adjusted R\textsuperscript{2} and significant predictors.

\begin{table}[h]
\begin{center}
\begin{tabular}{l c c c c c}
\hline
 & Institution A & Institution B & Institution C & Institution D & Institution E \\
\hline
(Intercept) & $195.001^{***}$ & $181.688^{***}$ & $571.074^{***}$ & $1257.600^{**}$ & $325.057^{***}$ \\
            & $(28.884)$      & $(36.117)$      & $(27.888)$      & $(238.894)$     & $(37.293)$      \\
TF          & $0.014$         & $0.005$         & $0.047^{***}$   & $0.009$         & $0.054^{***}$   \\
            & $(0.012)$       & $(0.004)$       & $(0.008)$       & $(0.022)$       & $(0.006)$       \\
Salaries    & $43.981$        & $184.052^{***}$ & $134.903^{***}$ & $57.533$        & $202.483^{***}$ \\
            & $(22.743)$      & $(13.179)$      & $(21.673)$      & $(73.679)$      & $(18.436)$      \\
\hline
R$^2$       & $0.429$         & $0.948$         & $0.934$         & $0.382$         & $0.955$         \\
Adj. R$^2$  & $0.334$         & $0.942$         & $0.925$         & $0.135$         & $0.951$         \\
Num. obs.   & $15$            & $21$            & $17$            & $8$             & $25$            \\
\hline
\multicolumn{6}{l}{\scriptsize{$^{***}p<0.001$; $^{**}p<0.01$; $^{*}p<0.05$}}
\end{tabular}
\caption{Production Function Models - Predicting High-Impact Publications}
\label{table:hpub_coefficients}
\end{center}
\end{table}

As with publications (Section \ref{sec:pub_prod}), this result with low adjusted R\textsuperscript{2} and non-significant coefficients for Institution D is likely due to the limited number of observations for it in the model (only 8 years). Production function results are expected to improve with a larger set of data points. Institution A again has an insignificant salaries predictor, due to its  investment/disinvestment cycles.

Table \ref{tab:hpubsummary} summarizes the results of the production function model on high-impact publications in real-world terms. For example, we can see that for Institution E, each additional 100 TeraFLOPs represents an annual increase of 4.61 high impact, Nature Index publications, and each \$100k investment in RCD salaries corresponds with an additional 21.12 high-impact publications annually.

\begin{table}[htbp]
  \centering

    \begin{tabular}{lccccc}
 & Institution A & Institution B & Institution C & Institution D & Institution E \\
    \textbf{100 TeraFLOPS} & 1.37  & 0.46  & 4.66  & 0.88  & 4.61 \\
    \textbf{\$100k Salaries} & 4.40  & 18.41 & 13.49 & 5.75  & 21.12 \\
    \textbf{Adj. R$^2$} & .429 & .942 & .925 & .135 & .951 \\
    \end{tabular}%
    \caption{Summary - High-Impact Publications Output}
  \label{tab:hpubsummary}%
\end{table}%

\subsection{Relative Importance of Coefficients}
\label{sec:relaimpo}
Using the \texttt{lmg} method implemented in the R package \texttt{relaimpo} \citep{Gromping2007}, we can calculate the proportion of the overall variance explained by each input in the model.  The "Combined" column in Table \ref{tab:combined_rela_impo} shows the relative importance of each input per output, for the production function model built from the combined dataset, as well as itemized by each institution.

For example,  we can see that for Institution C, Institution B, and Institution E, the production function model for HERD expenditures has similar proportions of variance explained by TeraFLOPs available on campus and salary costs, averaging 31\% and 59\%, respectively. These models have high adjusted R\textsuperscript{2} values between 0.925 and 0.955. The impact of salaries on Institution A's models is far less than other institutions, due to its investment/disinvestment cycles described in Section \ref{sec:localfactors}.

The relative importance of the combined model shows salary as approximately twice as important as TeraFLOPs of capacity as a predictor. As noted regarding Table \ref{table:comb_regs} and \ref{tab:combsummary}, the relative importance for the combined dataset for the entire population is much lower than the models for each institution, due to the limited breadth and statistical noise of the historical dataset.


\begin{table}[h]
  \centering

    \begin{tabular}{lrrrrrr}
   \hline
   \multicolumn{7}{l}{\textbf{Publications}} \\
   \hline
          & \multicolumn{1}{c}{\textbf{Inst. A}} & \multicolumn{1}{c}{\textbf{Inst. B}} & \multicolumn{1}{c}{\textbf{Inst. C}} & \multicolumn{1}{c}{\textbf{Inst. D}} & \multicolumn{1}{c}{\textbf{Inst. E}} & \multicolumn{1}{c}{\textbf{Combined}} \\
    \textbf{TeraFLOPS} & 0.070 & 0.260 & 0.402 & 0.508 & 0.345 & 0.135 \\
    \textbf{RCD Salaries} & 0.211 & 0.716 & 0.520 & 0.276 & 0.593 & 0.400 \\
    
    \hline
    \multicolumn{7}{l}{\textbf{Earned Doctorates}} \\
    \hline
          & \multicolumn{1}{c}{\textbf{Inst. A}} & \multicolumn{1}{c}{\textbf{Inst. B}} & \multicolumn{1}{c}{\textbf{Inst. C}} & \multicolumn{1}{c}{\textbf{Inst. D}} & \multicolumn{1}{c}{\textbf{Inst. E}} & \multicolumn{1}{c}{\textbf{Combined}} \\
    \textbf{TeraFLOPS} & 0..530 & 0.110 & 0.077 & 0.195 & 0.381 & 0.091 \\
    \textbf{RCD Salaries} & 0.237 & 0.589 & 0.540 & 0.043 & 0.560 & 0.250 \\
    
    \hline
    \multicolumn{7}{l}{\textbf{HERD Expenditures (\$M)}} \\
    \hline
          & \multicolumn{1}{c}{\textbf{Inst. A}} & \multicolumn{1}{c}{\textbf{Inst. B}} & \multicolumn{1}{c}{\textbf{Inst. C}} & \multicolumn{1}{c}{\textbf{Inst. D}} & \multicolumn{1}{c}{\textbf{Inst. E}} & \multicolumn{1}{c}{\textbf{Combined}} \\
    \textbf{TeraFLOPS} & 0.724 & 0.340 & 0.435 & 0.450 & 0.420 & 0.154 \\
    \textbf{RCD Salaries} & 0.093 & 0.565 & 0.487 & 0.536 & 0.517 & 0.403 \\
    
    \hline
    \multicolumn{7}{l}{\textbf{Hi Impact Publications}} \\
    \hline
          & \multicolumn{1}{c}{\textbf{Inst. A}} & \multicolumn{1}{c}{\textbf{Inst. B}} & \multicolumn{1}{c}{\textbf{Inst. C}} & \multicolumn{1}{c}{\textbf{Inst. D}} & \multicolumn{1}{c}{\textbf{Inst. E}} & \multicolumn{1}{c}{\textbf{Combined}} \\
    \textbf{TeraFLOPS} & 0.158 & 0.193 & 0.466 & 0.164 & 0.426 & 0.185 \\
    \textbf{RCD Salaries} & 0.270 & 0.755 & 0.468 & 0.219 & 0.529 & 0.450 \\
    
    \end{tabular}%
    \caption{Relative Importance, All Production Function Models}
  \label{tab:combined_rela_impo}%
\end{table}%


\subsection{Benchmark Models - Function of R\&D Expenditures}
\label{sec:func_of_herd}

Based on the 28 responses to the survey, the average ratio of TeraFLOPs of capacity to \$1M of R\&D expenditures is 11.47 TeraFLOPs per million dollars. (\textbf{Benchmark Q1}), and the average percentage of salary costs vs. R\&D Expenditures is 0.29\%  of R\&D expenditures (\textbf{Benchmark Q2}). 

With the value of the metrics of TeraFLOPs or salary costs $Investment$  shown as a function of institutional output R\&D expenditures $Investment = f(D_{herd})$, these coefficients are shown in Equations \ref{eq:tf_herd} and \ref{eq:sal_herd} (\textbf{Benchmark Q3}).

\begin{minipage}{0.45\textwidth}
\begin{equation} \label{eq:tf_herd}
{TF} = 11.47  \times D_{herd} 
\end{equation}
\end{minipage}
\begin{minipage}{0.45\textwidth}
\begin{equation} \label{eq:sal_herd}
{Salaries} = 0.29\%  \times D_{herd}
\end{equation}
\end{minipage}

Using these coefficients, Table \ref{tab:smith_model_results} presents the results of the benchmarking models, as a function of R\&D expenditures, based on the surveyed data in 2025. To compare these two models to reality:

By 2023, Institution E has exceeded \$845M of expenditures, with the benchmarking model calculating 9.7 petaFLOPs and \$2.5M of staffing for that level of expenditures. Institution E reported 10.9 of petaFLOPs of capacity, and \$2.65M of actual salary costs to the survey as of the winter of 2024-25.

\begin{table}[ht]

\begin{minipage}{.4\textwidth}

    \begin{tabular}{c|c}
    \multicolumn{1}{p{5em}}{{\centering 2023 R\&D Exp (\$M)}} & \multicolumn{1}{|p{6.5em}}{{\centering Modeled TF Capacity (2025)}} \\
    \hline \\
    \$1,900  &              21,784  \\
    \$1,500  &              17,198  \\
    \$1,200  &              13,758  \\
    \$1,000  &              11,465  \\
    \$850   &                 9,745  \\
    \$750   &                 8,599  \\
    \$400   &                 4,586  \\
    \$200   &                 2,293  \\
    \end{tabular}%

\end{minipage}
\hfil
\begin{minipage}{.55\textwidth}

   \begin{tabular}{c|c|c}

    \multicolumn{1}{p{5em}}{{2023 R\&D Exp (\$M)}} & 
    \multicolumn{1}{|p{6.2em}}{{\centering Modeled RCD Salaries (in 2025 \$M)}} & \multicolumn{1}{|p{6.2em}}{{\centering Modeled Total Budget (in 2025 \$M)}}\\
     \hline \\
    \$1,900  & \$5.59 & \$16.45 \\
    \$1,500  & \$4.42 & \$12.99 \\
    \$1,200  & \$3.53 & \$10.39 \\
    \$1,000  & \$2.94 & \$8.66 \\
    \$850   & \$2.50 & \$7.36 \\
    \$750   & \$2.21 & \$6.49 \\
    \$400   & \$1.18 & \$3.46 \\
    \$200   & \$0.59 & \$1.73 \\
    \end{tabular}%

\end{minipage}

        \caption{Center Investment Benchmarks - Model Results Based on TeraFLOPS}
  \label{tab:smith_model_results}%
\end{table}%

\subsubsection{Estimating Total Budgets}

 \cite{Lozado2023} for Gartner, Inc. report that ``34\% of enterprise IT spending is allocated to personnel salaries and benefits''. Using this formula and the known salary and wage expense at the institutions in the current investment survey dataset, we can estimate each respondent's total annual RCD budget, and subsequently compute the fraction of R\&D expenditures that budget represents. 

On average, research computing budgets of the 28 institutions in the current investment survey dataset are estimated at \$5.88M, or an average of 0.79\% of annual R\&D expenditures.

\subsection{Benchmark Models - Function of Earned Doctorates}
\label{sec:func_of_phd}
\begin{centering}
\begin{table}[h]

\begin{minipage}{.4\textwidth}

    \begin{tabular}{c|c}

    \multicolumn{1}{p{5em}}{\centering 2023 Earned Doctorates}& \multicolumn{1}{|p{6.5em}}{\centering Modeled TF (2025)}  \\
            \hline \\
    800   &              15,718  \\
    700   &              13,753  \\
    600   &              11,789  \\
    500   &                 9,824  \\
    400   &                 7,859  \\
    200   &                 3,929  \\
    \end{tabular}%

\end{minipage}
\hfil
\begin{minipage}{.5\textwidth}

    \begin{tabular}{c|c|c}

    \multicolumn{1}{p{5em}}{\centering 2023 Earned Doctorates} & 
    \multicolumn{1}{|p{6.5em}}{\centering Modeled RCD Salaries (in 2025 \$M)} &
    \multicolumn{1}{|p{6.2em}}{{\centering Modeled Total Budget (in 2025 \$M)}}\\
            \hline \\
    800   & \$3.76 & \$11.05\\
    700   & \$3.29 & \$9.67 \\
    600   & \$2.82 & \$8.29 \\
    500   & \$2.35 & \$6.91 \\
    400   & \$1.88 & \$5.52 \\
    200   & \$0.94 & \$2.76 \\
    \end{tabular}%

\end{minipage}

        \caption{Center Investment Benchmarks - Model Results Based on Earned Doctorates}
  \label{tab:stuff_per_phd}%
\end{table}%
\end{centering}

The average ratio of TeraFLOPs of capacity to earned doctorates is 19.65 TeraFLOPs per earned doctorate. (\textbf{Benchmark Q1}), and the average salary cost per earned doctorate is \$4,696 (\textbf{Benchmark Q2}).

With the value of the metrics of TeraFLOPs or salary costs $Investment$  shown as a function of institutional output earned doctorates $Investment = f(N_{phd})$, these coefficients are shown in Equations \ref{eq:tf_phd} and \ref{eq:sal_phd} (\textbf{Benchmark Q3}).

\begin{minipage}{0.45\textwidth}
\begin{equation} \label{eq:tf_phd}
{TF} = 19.65  \times N_{phd} \\
\end{equation}
\end{minipage}
\begin{minipage}{0.45\textwidth}
\begin{equation} \label{eq:sal_phd}
{Salaries} = \$4,696  \times N_{phd}
\end{equation}
\end{minipage}

Table \ref{tab:stuff_per_phd} shows the estimated TeraFLOPs, salary costs, and total RCD budgets needed by an institution at various quantities of PhD graduates, based on these coefficients.  To compare these two models to reality:

In 2023, Institution E awarded 810 doctoral degrees, with the benchmarking model calculating 14 petaFLOPs and \$3.8M of staffing for that number of graduates. Institution E reported 10.9 of petaFLOPs of capacity, and \$2.65M of actual salary costs to the survey, as of the winter of 2024-25.

\subsection{Benchmark Models - Function of Publications}
\label{sec:func_of_pubs}

Results of the historical dataset indicate that publications-based models are strong, therefore we present, as another option, a model publication data in the survey dataset.

The average ratio of TeraFLOPs of capacity to publication output is 1.34 TeraFLOPs per publication reported in Scopus, (\textbf{Benchmark Q1}), and the average salary cost per earned doctorate is \$3,400 (\textbf{Benchmark Q2}).

With the value of the metrics of TeraFLOPs or salary costs $Investment$  shown as a function of institutional output earned doctorates $Investment = f(N_{pub})$, these coefficients are shown in Equations \ref{eq:tf_pub} and \ref{eq:sal_pub} (\textbf{Benchmark Q3}).

\begin{minipage}{0.45\textwidth}
\begin{equation} \label{eq:tf_pub}
{TF} = 1.34  \times N_{pub} \\
\end{equation}
\end{minipage}
\begin{minipage}{0.45\textwidth}
\begin{equation} \label{eq:sal_pub}
{Salaries} = \$3,400  \times N_{pub}
\end{equation}
\end{minipage}

\begin{centering}
\begin{table}[h]

\begin{minipage}{.4\textwidth}

    \begin{tabular}{c|c}

    \multicolumn{1}{p{5em}}{\centering 2023 Publications}& \multicolumn{1}{|p{6.5em}}{\centering Modeled TF (2025)}  \\
            \hline \\
    20,000   &              26,744  \\
    14,000   &              18,721  \\
    10,000   &              13,372  \\
    6000   &                 8,023  \\
    3000   &                 4,011  \\
    1000   &                 1,337  \\
    \end{tabular}%

\end{minipage}
\hfil
\begin{minipage}{.5\textwidth}

    \begin{tabular}{c|c|c}

    \multicolumn{1}{p{5em}}{\centering 2023 Publications} & 
    \multicolumn{1}{|p{6.5em}}{\centering Modeled RCD Salaries (in 2025 \$M)} &
    \multicolumn{1}{|p{6.2em}}{{\centering Modeled Total Budget (in 2025 \$M)}}\\
            \hline \\
    20,000   & \$6.82 & \$20.07\\
    14,000   & \$4.78 & \$14.05 \\
    10,000   & \$3.41 & \$10.04 \\
    6000   & \$2.04 & \$6.02 \\
    3000   & \$1.02 & \$3.01 \\
    1000   & \$0.34 & \$1.00 \\
    \end{tabular}%

\end{minipage}

        \caption{Center Investment Benchmarks - Model Results Based on Publications}
  \label{tab:stuff_per_pub}%
\end{table}%
\end{centering}

Table \ref{tab:stuff_per_pub} shows the estimated TeraFLOPs, salary costs, and total RCD budgets needed by an institution at various quantities of publication output, based on these coefficients.  To compare these two models to reality:

In 2023, Institution E authors had 7,649 publications in the Scopus database, with the benchmarking model calculating 10.2 petaFLOPs and \$2.61M of staffing for that number of graduates. Institution E reported 10.9 of petaFLOPs of capacity, and \$2.65M of actual salary costs to the survey, as of the winter of 2024-25.

\section{Discussion}

With these results, we present a set of metrics and models that will allow university leaders to generalize the work by \cite{SmithMetrics2024} and validate this initial model for institutions beyond Purdue University, and apply the concept of supercomputing as a function of institutional outputs to capacity planning and budgeting. 

\begin{table}[h]
  \centering
 
    \begin{tabular}{lcc|cc}
          &        \multicolumn{2}{c}{\textbf{Historical Dataset}} & \multicolumn{2}{c}{\textbf{Smith, 2024}} \\
    \hline
    \textbf{Publications} &        \multicolumn{1}{l}{\textbf{Coefficients}} & \textbf{Rel. Imp.} & \multicolumn{1}{l}{\textbf{Coefficients}} & \textbf{Rel. Imp.} \\
    \hline
    \textbf{100 TeraFLOPS} &        19.59 & 14\%   & 7.14  & 29\% \\
    \textbf{\$100k RCD Salaries} &        136.90 & 40\% & 155.30 & 65\% \\
    \hline
    \textbf{Earned Doctorates} &              &       &       &  \\
    \hline
    \textbf{100 TeraFLOPS} &        1.24  & 9\%   & 2.55  & 31\% \\
    \textbf{\$100k RCD Salaries} &        7.83  & 25\%  & 7.36  & 42\% \\
    \hline
    \textbf{HERD Expenditures (\$M)} &              &       &       &  \\
    \hline
    \textbf{100 TeraFLOPS} &        2.99  & 15\%  & 2.59  & 25\% \\
    \textbf{\$100k RCD Salaries} &        14.46 & 40\%  & 9.04  & 43\% \\
    \hline
    \textbf{High Impact Publications} &              &       &       &  \\
    \hline
    \textbf{100 TeraFLOPS} &        3.59  & 19\%  & 3.34  & 35\% \\
    \textbf{\$100k RCD Salaries} &        21.21 & 45\%  & 22.98 & 61\% \\
    \end{tabular}%
  \label{tab:compare_to_smith}%
   \caption{Historical Dataset Summary - vs Smith (2024)}
\end{table}%

Although the historical dataset has a limited sample size (5 institutions, 8-24 years per institution), this study provides some confirmation that, with some exceptions such as earned doctorates, investment in either computing resources or staffing corresponds with increases of multiple institutional outputs. Investment in the RCD center's staff contributes for the largest relative importance in all models, as with Smith. Data from many more institutions, and over a longer time horizon, must be collected and analyzed in order to provide more conclusive results that fully reflect the general trends.

Specific models based on the current investment survey dataset are provided to give administrators a tool with which to size their investment based on the institution's outputs. On average, the 28 institutions in the current investment survey dataset operate 11.47 TeraFLOPS of central capacity per \$1M of R\&D Expenditures, and spend an amount equal to 0.29\% of their R\&D expenditures on research computing salaries. 

\subsection{Production Function Model Results}
\label{sec:model_discussion}

Table \ref{tab:all_results} summarizes the results of production function models from the historical dataset for all outputs, further separated by institution and for the entire dataset.  

\begin{table}[h]
  \centering

    \begin{tabular}{lrrrrrr}
   \hline
   \multicolumn{7}{l}{\textbf{Publications}} \\
   \hline
          & \multicolumn{1}{c}{\textbf{Inst. A}} & \multicolumn{1}{c}{\textbf{Inst. B}} & \multicolumn{1}{c}{\textbf{Inst. C}} & \multicolumn{1}{c}{\textbf{Inst. D}} & \multicolumn{1}{c}{\textbf{Inst. E}} & \multicolumn{1}{c}{\textbf{Combined}} \\
    \textbf{100 TeraFLOPS} & 40.41 & 8.18  & 21.57 & 23.09 & 23.12 & 19.59 \\
    \textbf{\$100k Salaries} & 0.04  & 113.60 & 87.04 & 14.92 & 136.9 & 136.50 \\
    \textbf{Adj. R$^2$} & .904 & .974 & .921 & .697 & .933 & .524\\

    \hline
    \multicolumn{7}{l}{\textbf{Earned Doctorates}} \\
    \hline
          & \multicolumn{1}{c}{\textbf{Inst. A}} & \multicolumn{1}{c}{\textbf{Inst. B}} & \multicolumn{1}{c}{\textbf{Inst. C}} & \multicolumn{1}{c}{\textbf{Inst. D}} & \multicolumn{1}{c}{\textbf{Inst. E}} & \multicolumn{1}{c}{\textbf{Combined}} \\
    \textbf{TeraFLOPS} & 2.62  & -0.05 & -0.26 & -2.05 & 2.10  & 1.24 \\
    \textbf{Salaries} & 2.38  & 4.74  & 3.88  & 3.04  & 9.86  & 7.83 \\
    \textbf{Adj. R$^2$} & .728 & .665 & .563 & -.067 & .936 & .325\\

    \hline
    \multicolumn{7}{l}{\textbf{HERD Expenditures (\$M)}} \\
    \hline
          & \multicolumn{1}{c}{\textbf{Inst. A}} & \multicolumn{1}{c}{\textbf{Inst. B}} & \multicolumn{1}{c}{\textbf{Inst. C}} & \multicolumn{1}{c}{\textbf{Inst. D}} & \multicolumn{1}{c}{\textbf{Inst. E}} & \multicolumn{1}{c}{\textbf{Combined}} \\
    \textbf{100 TeraFLOPS} & 4.64  & 2.13  & 3.38  & 2.73  & 3.18  & 2.99 \\
    \textbf{\$100k Salaries} & 0.31  & 13.10 & 11.23 & 13.63 & 11.85 & 14.46 \\
    \textbf{Adj. R$^2$} & .787 & .895 & .911 & .981 & .931 & .625 \\

    \hline
    \multicolumn{7}{l}{\textbf{Hi Impact Publications}} \\
    \hline
          & \multicolumn{1}{c}{\textbf{Inst. A}} & \multicolumn{1}{c}{\textbf{Inst. B}} & \multicolumn{1}{c}{\textbf{Inst. C}} & \multicolumn{1}{c}{\textbf{Inst. D}} & \multicolumn{1}{c}{\textbf{Inst. E}} & \multicolumn{1}{c}{\textbf{Combined}} \\
    \textbf{100 TeraFLOPS} & 1.37  & 0.46  & 4.66  & 0.88  & 5.37  & 3.59 \\
    \textbf{\$100k Salaries} & 4.40  & 18.41 & 13.49 & 5.75  & 20.25 & 21.21 \\
    \textbf{Adj. R$^2$} & .429 & .942 & .925 & .135 & .951 & .556\\

    \end{tabular}%
    \caption{Model Results - All Production Function Models}
  \label{tab:all_results}%
\end{table}%

For the production model across the entire population, however, the resulting models report moderately strong adjusted R\textsuperscript{2} values for each output, with statistically significant predictors. Further discussion of this combined model can be found in Section \ref{sec:prodfunc}.

Section \ref{sec:relaimpo} describes the relative importance of each predictor to each output. The relative importance of the combined model shows salary as approximately two to three times as important as TeraFLOPs of capacity as a predictor (Table \ref{tab:combined_rela_impo}).

  Most output metrics show still strong (> 50\%) average relative importance of their capital and labor predictors \cite[as in][]{SmithMetrics2024}, at 60\%/30\%. Table \ref{tab:combined_rela_impo} further reports the specific relative importance of each predictor to each output for each individual institution.

\subsubsection{Local Factors Influencing Quality of Models}
\label{sec:localfactors}
External factors with Institution A's and Institution D's salary investment and earned doctorates, respectively, limit the quality and utility of several models. 

\begin{figure}[h]
\centering
\subfloat[Institution D Earned Doctorates 2000-2022]{
    \centering
    \includegraphics[width=.45\textwidth]{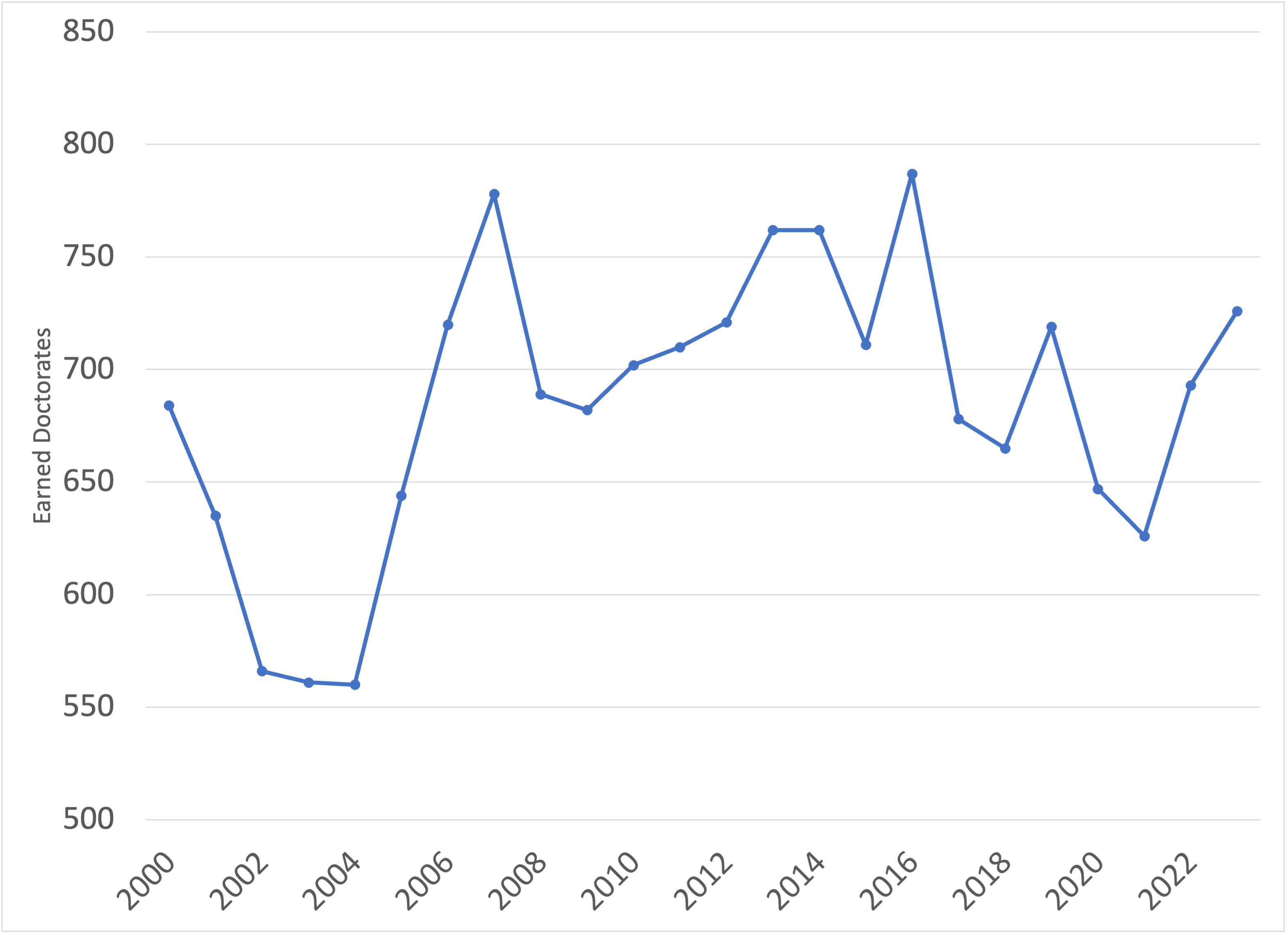}
    \label{fig:umn_phds}
}
\subfloat[Institution A RCD Salary Costs 2009-2022]{
    \centering
    \includegraphics[width=.45\textwidth]{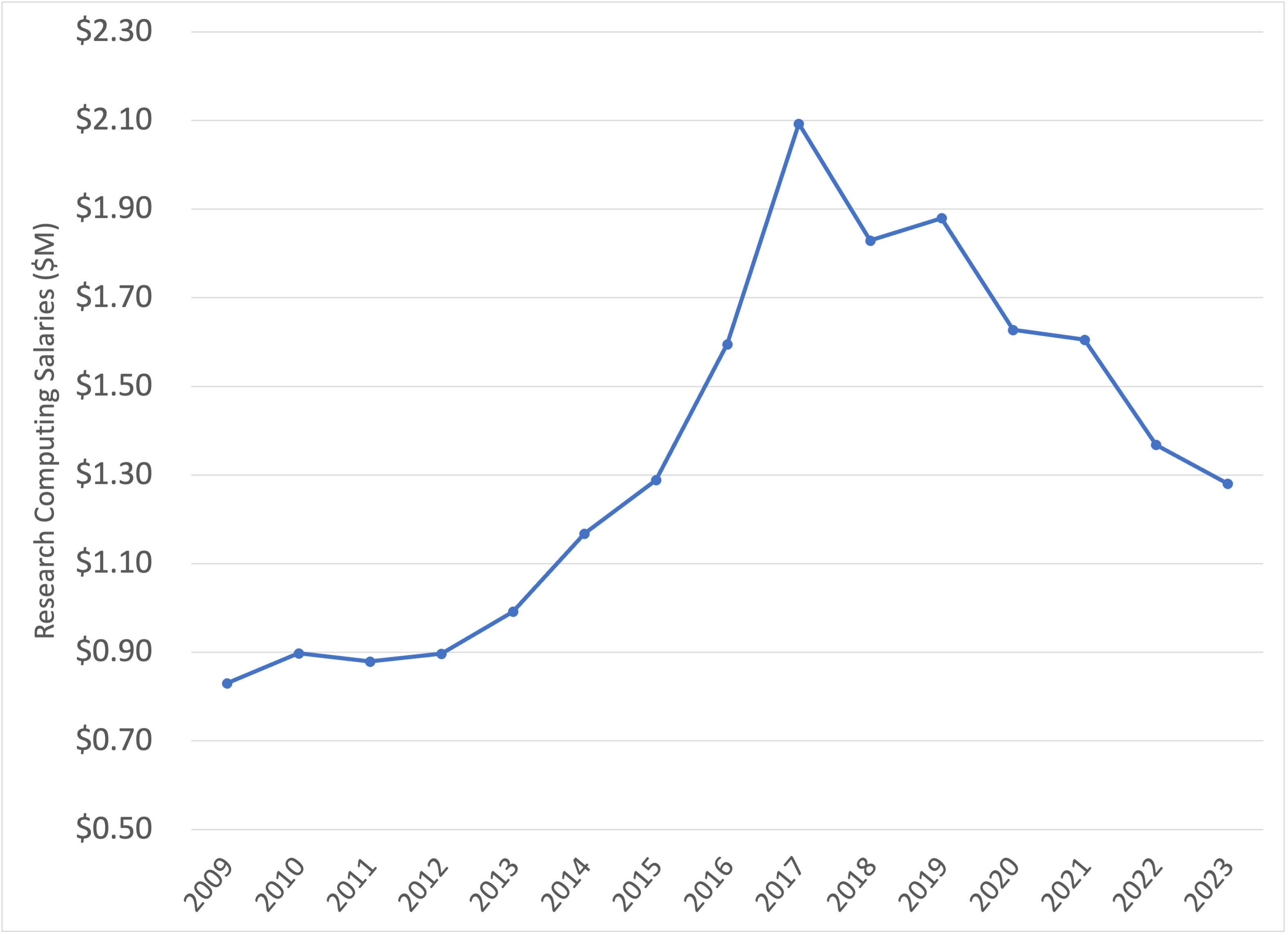} 
    \label{fig:clemson_salaries}
}

\caption[short]{Outside Variability in Source Data.}
\end{figure}

Observed both in correlation analysis and several production function models, the weakness of earned doctorates in Institution D's models is likely due to a fluctuating trend in doctoral degrees awarded there (Figure \ref{fig:umn_phds}), rising in the early 2000s, followed by a highly variable period over the last decade, generally declining. 

Furthermore, the overall weak results predicting earned doctorates at multiple institutions may indicate that \cite{SmithMetrics2024} studying Purdue may in fact be for an atypical institution in this set.  Purdue's output in earned doctorates has steadily climbed since 2003, rising to the 3rd largest producer of PhDs in the 2023 Survey of Earned Doctorates, with only one other institution in the historical dataset ranking in the top 25. If data were available, performing this analysis for other very large producers of PhDs such as Michigan, Stanford, or Texas A\&M to reproduce the results from Smith would be informative.

Section \ref{sec:future_work_prod} describes future enhancements to the models that will mitigate effects such as these.

While still positive, Institution A shows weaker correlations than other institutions due to variability in its RCD salary investment over time (Figure \ref{fig:clemson_salaries}). This variability is attributed to shifting institutional priorities. RCD salary investment grew steadily until the 2017 retirement of the Provost, the President, and a research-focused CIO. This significant change in leadership was followed by a period of steady budget reductions until the current CIO, hired in 2021, reversed the decline. This external factor suggests that salary investment for Institution A may be a weaker predictor than was observed in previous work, or with other institutions in this study.

\subsection{Applications of benchmarking as a function of output}

With outputs and the coefficients in the equations $Investment = f(D_{output})$ identified, there are several situations where a university leader might apply them. 

\subsubsection{Selecting an Output as a  Driver of Decision-Making}
Benchmarking models based on R\&D expenditures, earned doctorates, or publications present slightly different pictures for the same institution. 

Local factors certainly do come into play: consider universities with very large research expenditures but relatively low numbers of earned doctorates (Columbia), or ones with extremely large numbers of each (Michigan). The ideal value for a given university may lie at a value in between results provided by each model.

\begin{figure}[h]
\setcounter{subfigure}{0}
\centering
\subfloat[TeraFLOPs]{
  \centering
    \includegraphics[width=.45\textwidth]{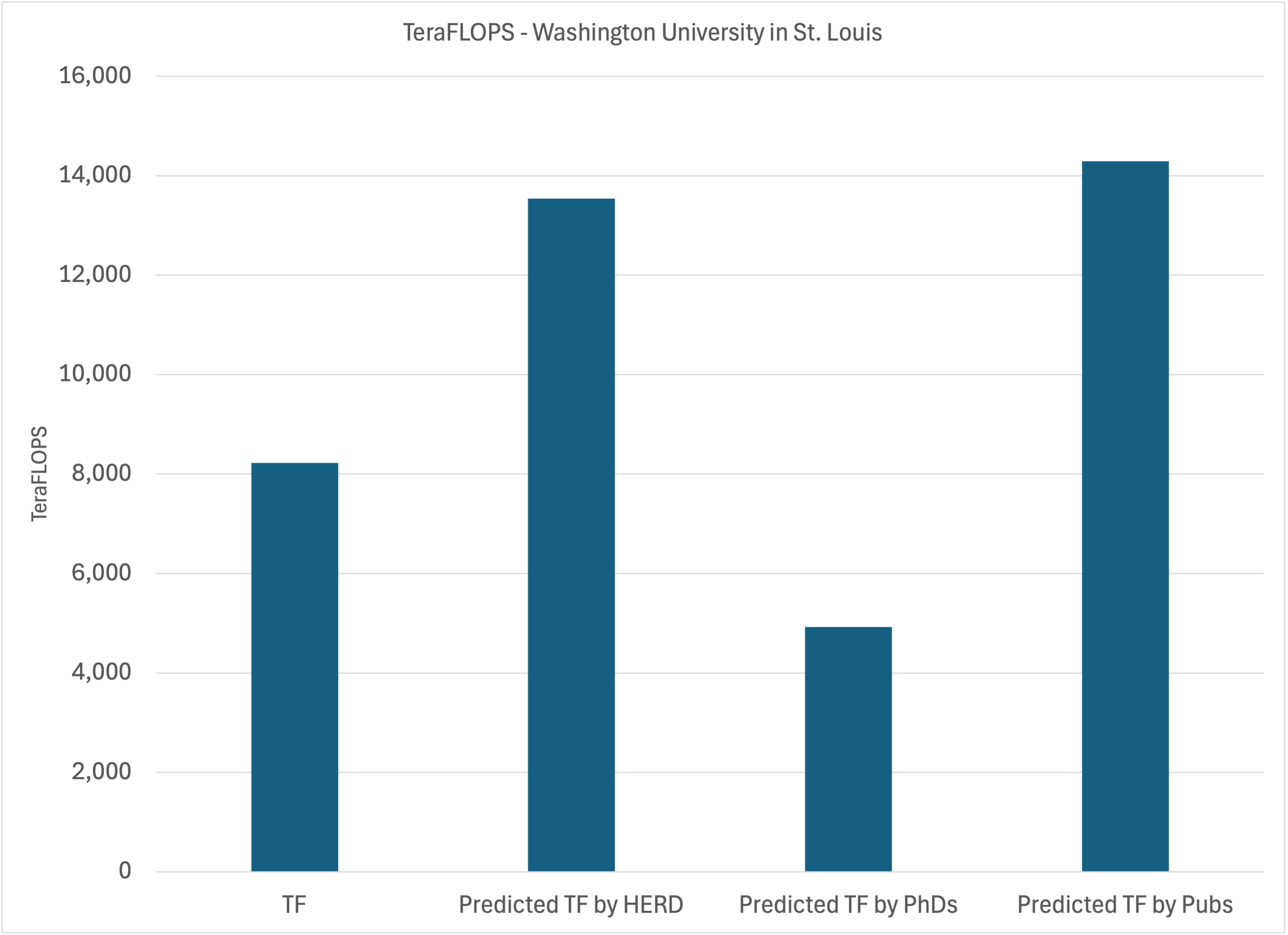} 
    \label{fig:wustl_tf}
}
\subfloat[Salaries]{
  \centering
    \includegraphics[width=.45\textwidth]{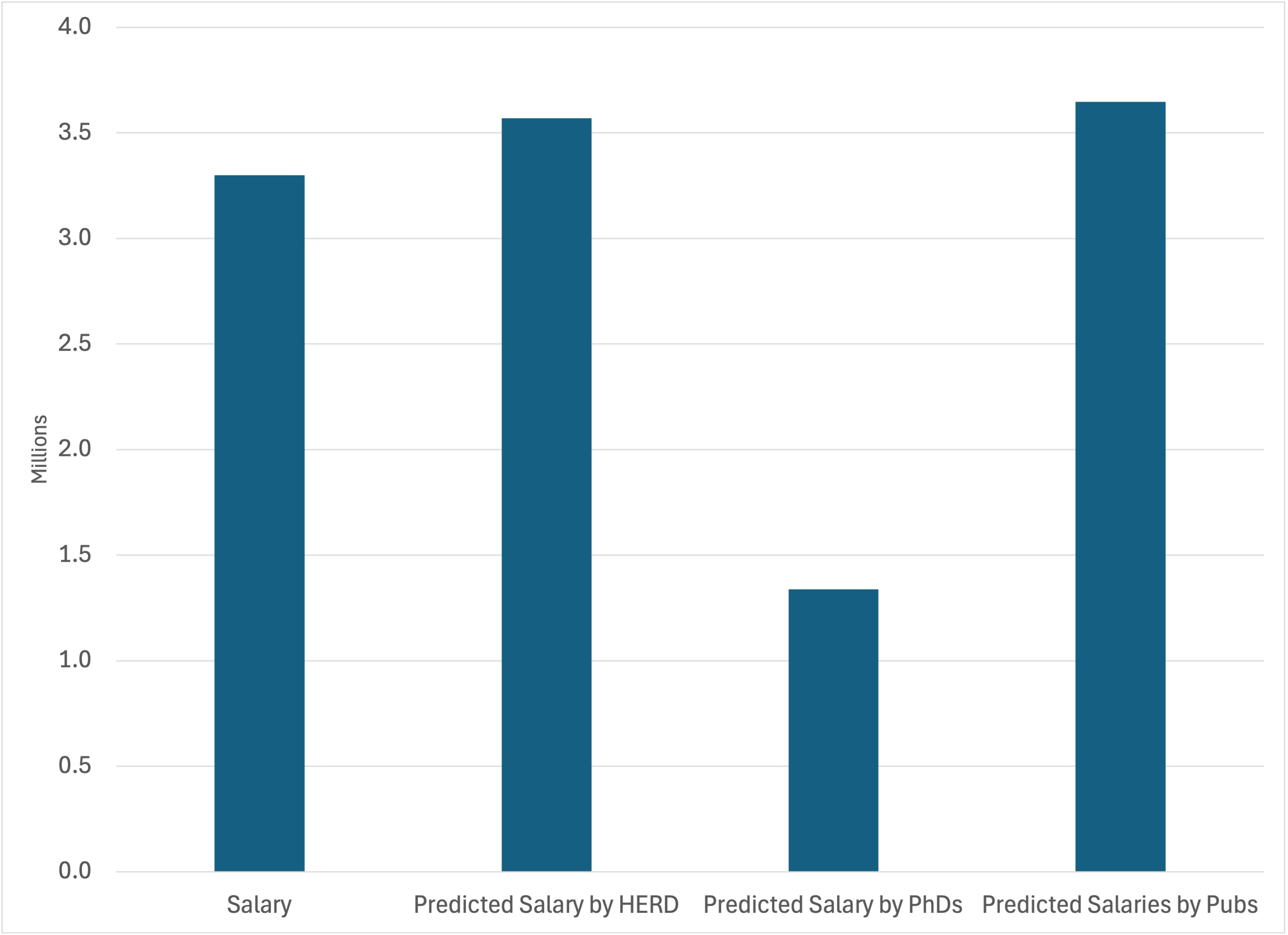} 
    \label{fig:wustl_salaries}
}
\caption[short]{Example Institution - Models Predicting Scales}
\end{figure}

Figures \ref{fig:wustl_tf} and \ref{fig:wustl_salaries} show modeled vs actual data for an example survey respondent, an R1 from the central United States. In the case of this institution, their actual capacity in TeraFLOPS is almost right in between the values predicted by expenditures and earned doctorates, while their salary investment aligns much closer to the model based on HERD expenditures. 

In both examples, the results from the model based on publications is very near to the one based on R\&D expenditures. In the historical dataset, publications and R\&D expenditures are highly correlated ($\tau_{pub} = .985, p < .05$). R\&D expenditures are a leading indicator of future publications and may be a better practical metric to drive decision-making.

\subsubsection{Institutional Benchmarking}
\begin{figure}[h]
\setcounter{subfigure}{0}
\centering
\subfloat[Based on HERD Expenditures]{
  \centering
    \includegraphics[width=.45\textwidth]{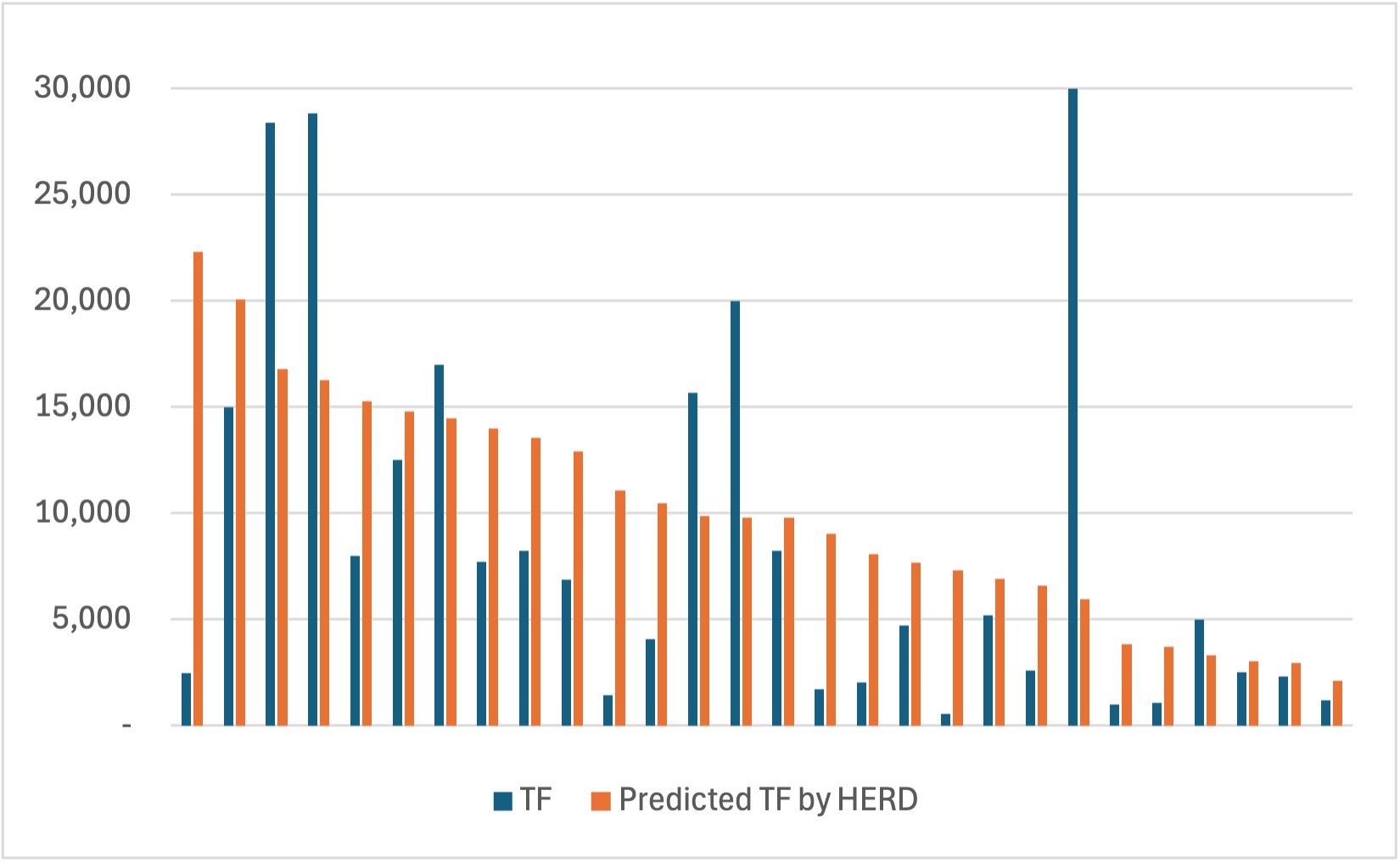} 
    \label{fig:tf_all_herd}
}
\subfloat[Based on Earned Doctorates]{
  \centering
    \includegraphics[width=.45\textwidth]{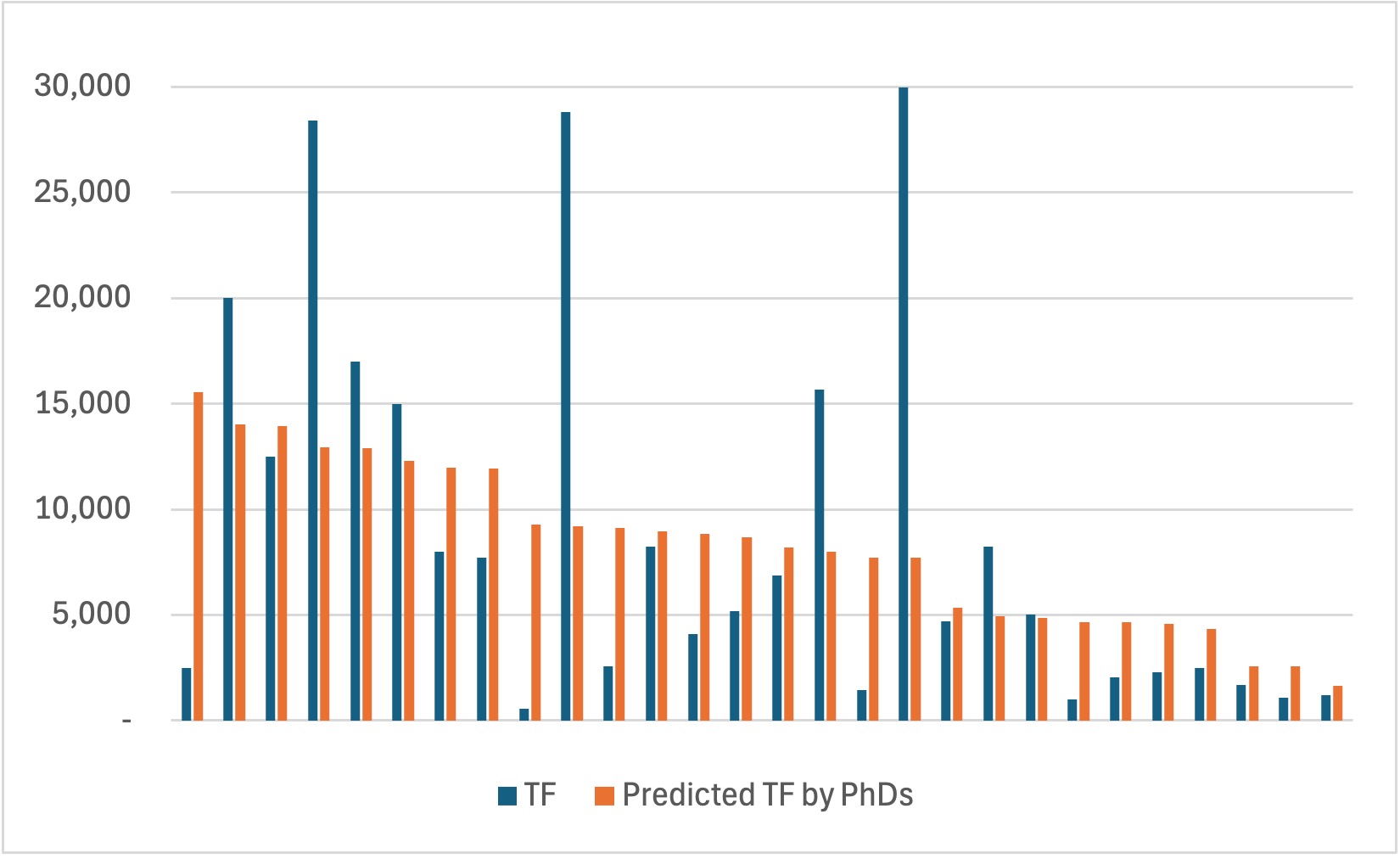} 
    \label{fig:tf_all_phd}
}
\caption[short]{Predicted vs Actual TeraFLOPS - HERD and Doctorates Models}
\end{figure}

University administrators have finite resources and may ask themselves, "what size of an HPC operation does my institution need?" Under-resourcing risks impacting competitiveness, and over-resourcing is wasteful. Although the model, with only 28 respondents, may not be sufficiently fine-grained in the middle for subtle adjustments, it may already be useful at either end of the spectrum. Figures \ref{fig:tf_all_herd} and \ref{fig:tf_all_phd} show actual vs predicted TeraFLOPS for all 28 respondents, predicted by HERD expenditures and by earned doctorates. In these figures we see many institutions' actual capacity near their predicted value. Some outliers exist, however - both high and low -  suggesting that some institutions may not be correctly resourced relative to their output.

\subsubsection{Future Capacity Planning}

The 8 to 24 years of historical data used to inform the benchmarking models in \ref{sec:hist_dataset} provide us with an observation of historical trends across these five institutions, specifically showing an annual growth trend of 41\% in operated TeraFLOPS, rather closely tracking Moore's Law improvements.

\begin{figure}[htbp]
  \centering
    \includegraphics[width=.7\textwidth]{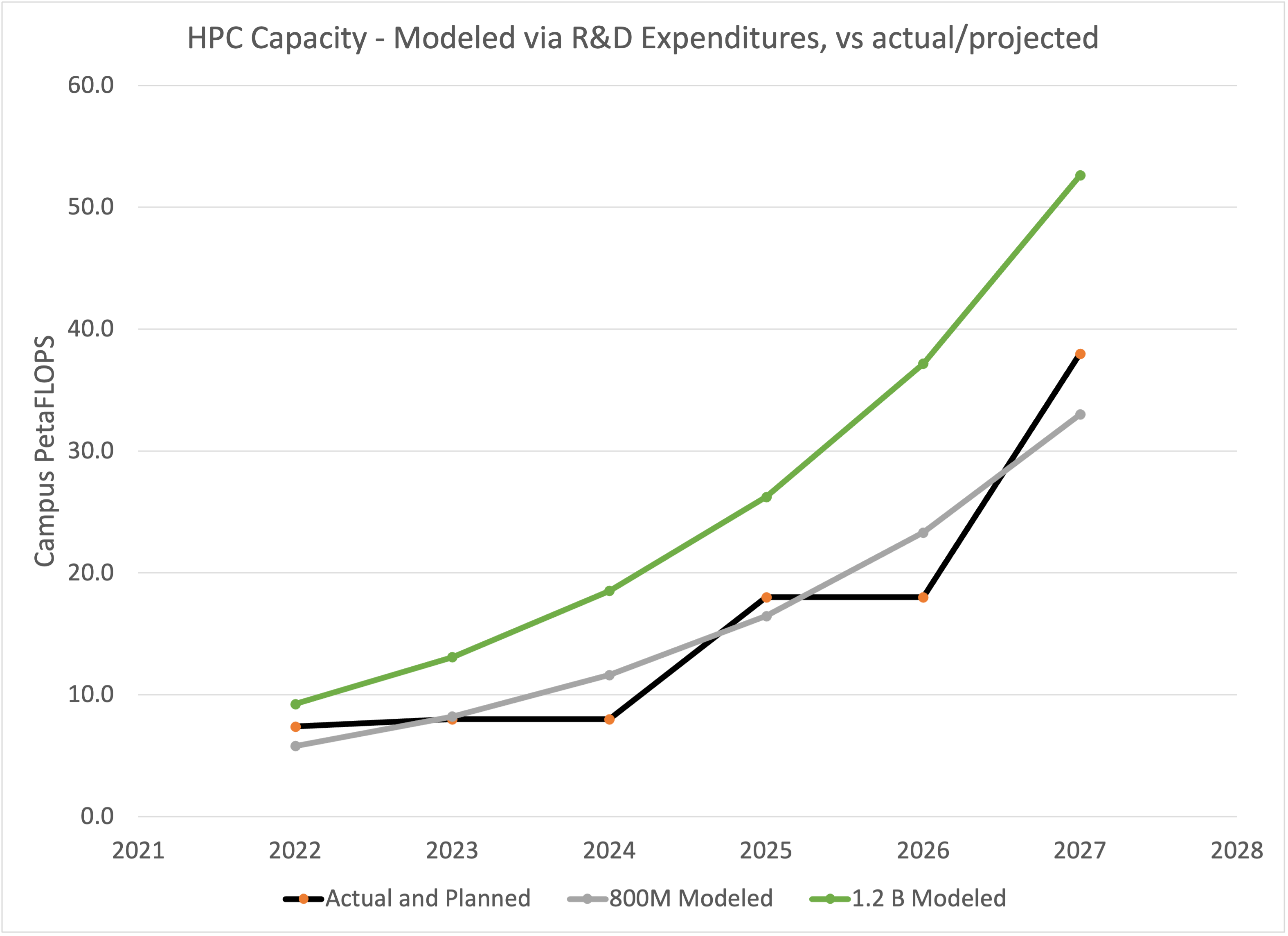} 
    \caption{\centering{Projecting Future Capacity Needs as a Function of R\&D Expenditures}} 
    \label{fig:purdue_curve}
\end{figure}

With this year-over-year growth, Figure \ref{fig:purdue_curve} shows future modeled capacity at \$800M and \$1.2B expenditure levels, common output levels for many major universities. Figure \ref{fig:purdue_curve} also shows the real deployed capacity at one university, including the early 2025 deployment of a new 10 PF system, and projections for future deployments based on long-term budget planning.

A university leader may use this model - or one based on future earned doctorates -  to plan future investments. For example, if this university were to be a billion dollar research enterprise in 2028, the model suggests that it may need 43-61 PetaFLOPs of capacity. 

In the real world, a \$1.2B university deploying a 50 PetaFLOP system in 2025 may look like an extremely large investment today, but looking ahead to 2027, such a system places them right in line with their expected 2027 point on the curve, in advance.

\subsection{Future Work}
\subsubsection{Production Function Model}
\label{sec:future_work_prod}
As discussed in Section \ref{sec:localfactors}, the external factors of declining and shifting trends in the doctoral degrees awarded at Institution D complicated modeling the earned doctorate output for that institution. Future work may explore a hypothesis that a university's largest fields in terms of PhDs awarded are likely the CI users, and refine the earned doctorate models to use ``earned doctorates from the institution's top X\% of fields of science'' as an output instead of the entire number of PhDs awarded.

Similarly, the boom-and-bust investment/disinvestment cycles at Institution A complicate attempts to model the relationship between investment and output. While a counter-example of an institution with a mixed track record for maintaining its investment in research computing is useful to have, as part of a small population its presence makes model outputs in this study inconclusive at best.

Finally, it stands to reason that there is a point of diminishing returns for institutional investment in one of the factors examined, with there being a point where an even bigger supercomputer stops adding additional value. Future work could potentially build on the interaction analysis touched upon by \cite{SmithDisseratation} and analyze the interactions between people and capital investments. This would address questions that can more precisely guide institutional decision-making -- for example, it could answer the question of whether the rate of impact on HERD expenditures from investing more in computational resources is different at higher or lower levels of investment in RCD center staff.  

\subsubsection{Benchmarking Model}
First, a benchmarking and scaling model may have a complex set of factors than simply R\&D expenditures or earned doctorates. Future work may identify a better-tuned model potentially combining both of these factors, or adding new ones such as undergraduate populations, the mix of research at an institution, public/private status, among others.

Next, the rapid growth in AI compute will require more nuance in the capacity metric used herein - mixed or lower-precision FLOPS will soon far outpace FP64 at university centers, and the model will need to evolve to reflect that.

Further exploration and breakdowns of institutional characteristics can be incorporated, such as the presence of a medical school or status as a land grant institution.

Finally, in the future, we aim to continue to perform this survey annually, with the goal of adding institutions and building a historical corpus of data with which to establish longer trends. This ongoing data collection would create a dataset suitable for statistically powerful production function models across many institutions.

\section{Acknowledgements}
This material is based upon work supported by the National Science Foundation under award numbers 2005632 and 2321090 (Smith), 2227627 (Snapp-Childs), 1445604 and 2005506 (Hancock).  Elements of this work are based on the doctoral dissertation by P. Smith \citep{SmithDisseratation}. 

 \clearpage
\bibliographystyle{acm}
\bibliography{anonymized}

\end{document}